# Gauge Theory in $d = 2 + 1$ at High Temperature: $Z_N$ interface[*]

C. Korthals Altes[a], A. Michels[b], M. Stephanov[c][†] and M. Teper[b]

[a]Centre Physique Theorique au CNRS, Luminy, B.P. 907, 13288 Marseille, France

[b]Theoretical Physics, U. of Oxford, 1 Keble Rd., Oxford OX1 3NP, UK

[c]Dept. of Physics, U. of Illinois at Urbana–Champaign, 1110 W. Green St., Urbana 61801–3080, USA

We calculate on the lattice the interface tension in the $SU(2)$ pure gauge theory in $d = 2+1$ at high temperature. The result is compared to the perturbative prediction. The agreement confirms applicability of the perturbation theory in this case.

## 1. Hot QCD and $Z_N$ symmetry

Studies of QCD at high temperature attracted considerable attention recently (see, e.g., [1,2] and refs. therein). One of the interesting questions concerns the existence and physical significance of the so–called $Z_N$ ($N = 3$) phases of quark–gluon plasma at high temperature ($T > T_c$). These phases appear naturally in Euclidean formulation of QCD. In such a formulation the system lives in a 4–dimensional box. The extent of the three spatial dimensions is large compared to the physical scale and the extent of the fourth (Euclidean time) dimension is $1/T$. The boundary conditions in this time direction are periodic (for gluon fields).

Consider a hypothetical QCD without quarks and generalize it to an $SU(N)$ gauge theory in $d$ space–time dimensions. An order parameter distinguishing between the $Z_N$ phases is the value of the Polyakov loop: a path ordered exponent along a path in the time direction at fixed spatial coordinate $\boldsymbol{x}$:

$$P(\boldsymbol{x}) = \frac{1}{N} \operatorname{Tr} P \exp\left\{ ig \int_0^{1/T} dt\, A_0(t, \boldsymbol{x}) \right\}, \quad (1)$$

where $A_\mu = A_\mu^a T^a$ is the $N \times N$ matrix of the gauge potential. It is gauge invariant due to periodic boundary conditions provided that the gauge transformation $U(\boldsymbol{x}, t)$ is also periodic in $t$. However, there are also gauge transformations which

---
[*]Contribution to the Lattice 94 conference, Sep. 27–Oct. 1, Bielefeld, Germany
[†]Supported by the grant NSF–PHY 92-00148.

are periodic up to an element of the center of the gauge group:

$$U(\boldsymbol{x}, t + 1/T) = zU(\boldsymbol{x}, t), \quad (2)$$

where $z = I \exp(i2\pi k/N)$, $k = 0, 1, 2, \ldots, N - 1$; $I$ is a unit $N \times N$ matrix. The periodic b.c. are not affected by such a gauge transformation (if there are no fermions) as $z$ commutes with all matrices in the gauge group. However, the Polyakov loop acquires a phase $\exp(2i\pi k/N)$.

If one integrates out all degrees of freedom except for the Polyakov loops the resulting effective theory of the complex scalar field $P(\boldsymbol{x})$ in $d-1$ dimensions will have a global $Z_N$ symmetry: $P \to \exp(i2\pi k/N)P$. It turns out [3] that at *high* $T$ this system is ferromagnetically ordered: $\langle P \rangle \neq 0$ (because an effective ferromagnetic coupling grows with $T$). The value of $\langle P \rangle$ can be related to the free energy $F$ of a static quark inserted in the hot gluon plasma: $\langle P \rangle = \exp(-F/T)$. Thus confinement corresponds to $\langle P \rangle = 0$, i.e., $F = \infty$. In this way one can relate the breaking of the $Z_N$ symmetry to the deconfinement transition.

## 2. Interfaces and perturbation theory

There are $N$ possible complex directions where the vacuum expectation value $\langle P \rangle$ can point. If two parts of space are occupied by domains with different values of $\langle P \rangle$ an interface occurs between the domains. Such interfaces might have interesting cosmological consequences [4].

Perturbation theory can be used to calculate properties of such interfaces at very high tem-perature when the effective gauge coupling becomes small. Integrating out quadratic fluctuations around constant homogeneous configuration $A_0$ one obtains the effective potential as a function of $A_0$. Consider, for example, an $SU(2)$ gauge theory. One can rotate the constant field $A_0$ to some direction in the matrix space, say, $A_0 = A_0^3 \tau_3/2$. Then the effective potential has the form as shown in Fig. 1 [5].

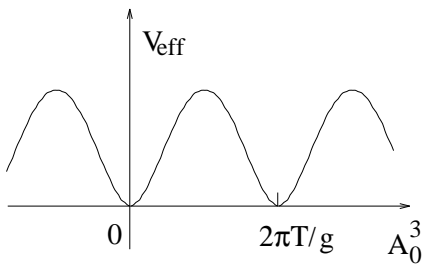

Figure 1. $V_{\text{eff}}$.

The periodicity of $V_{\text{eff}}$ reflects the $Z_2$ symmetry of the effective theory of the Polyakov loops: $P \to -P$. Indeed, for our constant $A_0$: $P = \cos(gA_0^3/2T)$. At high $T$ the $Z_2$ symmetry will be spontaneously broken: minima of $V_{\text{eff}}$ correspond to $P = \pm 1$.

One can calculate the interface tension between two phases by considering a soliton-like solution which starts in one minimum, goes through the barrier and ends in another minimum. The action on such a trajectory would give the interface tension. In $d = 4$ one obtains [6]: $\sigma = cT^3/g$, where $c$ is a numerical constant. This result is easy to understand: the width of the interface is of order $1/gT$ – Debye screening length, and the energy density inside the interface is of order $T^4$.

One has to take perturbative results at high temperature with caution, however. Corrections from higher orders of the perturbative expansion contain infrared divergencies. For example, it was suggested in [7] that infrared divergencies might result in vanishing interface tension.

The aim of this paper is to study a pure gauge theory nonperturbatively (on the lattice), determine the interface tension at very high temperatures ($T \gg T_c$) and compare these results with perturbation theory.

## 3. Lattice study

We study $SU(2)$ pure gauge theory in $d = 2+1$. The gauge variables — $SU(2)$ matrices — are defined on the links of a simple cubic $L_x \times L_y \times L_t$ lattice. The action is a sum over the plaquette products $\Box_P$ of the gauge matrices:

$$S = \sum_P (1 - \frac{1}{2} \operatorname{Tr} \Box_P) \qquad (3)$$

The partition function is $Z = \exp(-\beta S)$.

To enforce an interface we use a twist [8]. The action of a system with the twist differs from the untwisted action in that the contribution of a set of plaquettes pierced by a line in $y$ direction changes sign. One can check that this is equivalent to antiperiodic b.c. in $x$ direction on the Polyakov loops.

We simulate the system with and without the twist and measure the average action in the two cases. The difference $\Delta S$ is related to the excess free energy $\Delta F \equiv -\ln(Z_{\text{tw}}/Z)$ due to the interface and hence to the interface tension $\sigma$:

$$\Delta S = \frac{\partial \Delta F}{\partial \beta} = \frac{\partial}{\partial \beta} \left( \frac{\sigma A}{T} \right), \qquad (4)$$

where $A = L_y a$ is the "area" of the interface, $T = 1/(L_t a)$ is the temperature in physical units and $a$ is the lattice spacing.

We calculated the interface tension for this model in continuum to the leading order in $g^2$:

$$\sigma = \alpha_0 \frac{T^{5/2}}{g}, \qquad \alpha_0 = 5.104\ldots. \qquad (5)$$

The interface width is of order $g\sqrt{T}$. The $g^2$ is the (dimensionful) bare coupling constant for the $SU(2)$ gauge theory in continuum. It is related to the lattice parameter $\beta$ as: $\beta = 4/(g^2 a)$. The dimensionless small expansion parameter of the perturbation theory is $g^2/T$, which in terms of the lattice parameters is: $4L_t/\beta$.

We can now substitute (5) into (4) and express $A$, $T$ and $g$ through the lattice parameters $L_y$, $L_t$ and $\beta$ to get the leading large $\beta$ behavior of $\Delta S$ in perturbation theory:

$$\Delta S = \alpha_0 \frac{L_y}{4 L_t^{3/2}} \frac{1}{\sqrt{\beta}}. \qquad (6)$$

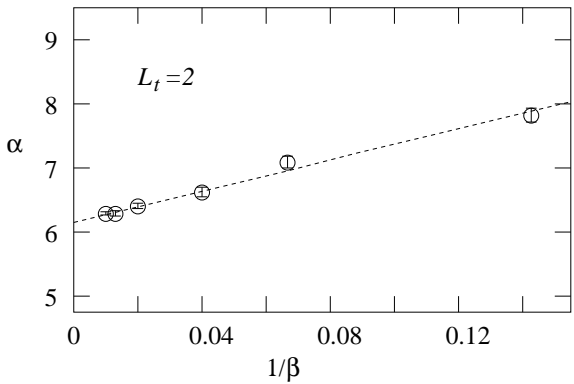

Figure 2. The value of $\alpha \equiv 4L_t^{3/2}\Delta S\sqrt{\beta}/L_y$ as a function of $1/\beta$ at a given $L_t$ and a linear extrapolation to $\beta = \infty$. (For comparison: $1/\beta_c \approx 0.29$ [9]).

We perform measurements at different values of $\beta$ at given $L_t = 2, 3, 4$. The spatial sizes $L_x$, $L_y$ are chosen after test runs to study finite size dependence. For $L_t = 2$ and $\beta = 15$ we chose $L_x = 36$ and $L_y = 12$ and then scaled the sizes with the interface width (roughly as $\sqrt{L_t\beta}$). We plot $4L_t^{3/2}\Delta S\sqrt{\beta}/L_y$ as a function of $1/\beta$ and extrapolate to $\beta = \infty$ (see, e.g., Fig. 2). The resulting $\alpha_0$ depends on $L_t$ which is the cutoff $1/a$ in units of the temperature $T$. The continuum limit is $L_t \to \infty$. It is possible, however, to calculate $\sigma$ to the leading order in $g^2/T \sim 1/\beta$ in perturbation theory for *finite* $L_t$. We compare our results to perturbation theory at each $L_t$ in Fig. 3.

The extrapolation to $1/\beta = 0$ (i.e., $T = \infty$) in Fig. 2 has an ambiguity as we do not know the form of the perturbative correction to $\sigma$ in (5). The linear $1/\beta$ extrapolation we used assumes this correction to be $O(gT^{3/2})$. While we do not know its precise form, a simple analysis shows that this correction could contain additional powers of[3] $\ln T/g^2 \sim \ln \beta$ which might also build up to a power of $\beta$. Such an ambiguity is of the same order as the deviation of the $1/\beta$ extrapolation from the leading perturbative result in Fig. 3. We checked this extrapolating with $1/\beta^{1-\epsilon}$, $0 < \epsilon < 0.5$.[4]

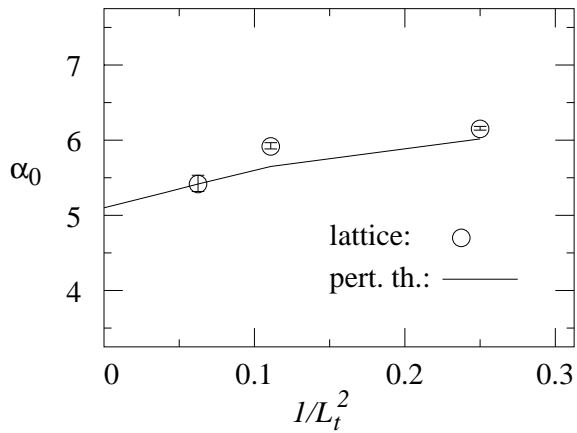

Figure 3. Comparison of the lattice results from the linear $1/\beta$ extrapolation to $\beta = \infty$ at fixed $L_t$ and from perturbation theory for $\alpha_0$ in (5), (6).

In conclusion, while our work is at a preliminary stage both analytically and numerically, we can already see that perturbation theory does indeed describe the properties of the interface very accurately.

---

[3] Such logarithms are due to the IR divergencies specific to d=2+1 and appear, e.g., in the calculation of the Debye mass in this theory [10].

[4] This results in smaller extrapolated $\alpha_0$ values.